 \documentclass[authoryear,preprint,review,12pt]{elsarticle}

\usepackage{color}

\usepackage{graphicx}
\usepackage{lineno}  
\usepackage{amssymb}
\usepackage{float}
\usepackage{arydshln}
\usepackage{url}

\usepackage{multirow}
\usepackage{threeparttable}
\usepackage{bm}
\usepackage{amsmath}

\usepackage[x11names]{xcolor}
\usepackage[
  colorlinks,
]{hyperref}

\AtBeginDocument{\hypersetup{citecolor=DodgerBlue4}}

\journal{}

\begin{document}

\begin{frontmatter}
\title{Feature-free regression kriging}


\author[a]{Peng Luo}
\ead{pengluo@mit.edu}

\author[b]{Yilong Wu}
\ead{22074689@student.curtin.edu.au}

\author[b]{Yongze Song}
\cortext[cor1]{Corresponding author: Yongze Song, yongze.song@curtin.edu.au}

\address[a]{Senseable City Lab, Massachusetts Institute of Technology, Cambridge, USA}

\address[b]{School of Design and the Built Environment, Curtin University, Perth, Australia}

\begin{abstract}

Spatial interpolation is a crucial task in geography. As perhaps the most widely used interpolation methods, geostatistical models—such as Ordinary Kriging (OK)—assume spatial stationarity, which makes it difficult to capture the nonstationary characteristics of geographic variables. A common solution is trend surface modeling (e.g., Regression Kriging, RK), which relies on external explanatory variables to model the trend and then applies geostatistical interpolation to the residuals. However, this approach requires high-quality and readily available explanatory variables, which are often lacking in many spatial interpolation scenarios—such as estimating heavy metal concentrations underground.
This study proposes a Feature-Free Regression Kriging (FFRK) method, which automatically extracts geospatial features—including local dependence, local heterogeneity, and geosimilarity—to construct a regression-based trend surface without requiring external explanatory variables. 
We conducted experiments on the spatial distribution prediction of three heavy metals in a mining area in Australia. In comparison with 17 classical interpolation methods, the results indicate that FFRK, which does not incorporate any explanatory variables and relies solely on extracted geospatial features, consistently outperforms both conventional Kriging techniques and machine learning models that depend on explanatory variables.
This approach effectively addresses spatial nonstationarity while reducing the cost of acquiring explanatory variables, improving both prediction accuracy and generalization ability.  This finding suggests that an accurate characterization of geospatial features based on domain knowledge can significantly enhance spatial prediction performance—potentially yielding greater improvements than merely adopting more advanced statistical models.

\end{abstract}

\end{frontmatter}


\section{Introduction}

Spatial interpolation is one of the essential tasks in geography \citep{lam1983spatial,goodchild2004giscience}. Based on samples collected from the Earth's surface (e.g., land and ocean), a continuous surface is estimated by appropriately modeling the relationships between the samples \citep{webster2007geostatistics}. With the new Earth observation technologies, such as remote sensing, have emerged in recent years, the availability of many geographical variables has increased dramatically \citep{campbell2011introduction}. However, there are still scenarios where the distribution of variables can only be obtained through sampling, such as the  soil organic matter \citep{cheng2024explainable} or the elemental content under the ocean \citep{luo2023generalized}. Recently, the potential of spatial interpolation method in deep space exploration, such as Mars mineral prediction, has been recognized \citep{jiao2025modeling}. In such cases, spatial interpolation remains a necessary step.

From the perspective of modern spatial statistics, the earliest interpolation methods are considered to belong to the category of deterministic interpolation, such as Inverse Distance Weighting (IDW) \citep{panigrahi2021inverse}. They based on an assumption that the value at a certain location can be obtained by inversely weighting the distances from surrounding observed points \citep{tomczak1998spatial}. However, this approach neglects spatial dependencies and fails to adequately consider the overall spatial distribution characteristics and patterns of observed points, limiting its interpolation accuracy. Geostatistical models were developed with a thorough consideration of spatial dependence \citep{matheron1963principles}. For example, the original geostatistical method was Ordinary Kriging \citep{cressie1988spatial}. They interpolate based on the spatial variation patterns in the data, resulting in more accurate and unbiased estimation results. This approach has found extensive applications across various fields — from geology and mining, where it is used to predict unknown geological features or the distribution of mineral deposits, to agriculture, environmental science, and ecology. The basic assumption of geostatistical models is second-order stationarity, which assumes that the variance between two samples depends only on their distance and direction, regardless of their absolute locations \citep{clark1979practical}.

However, the distribution of geographic variables can exhibit non-stationary characteristics \citep{cantrell1991effects}. For example, when a study area contains multiple distinct terrain structures, the distribution patterns of geographic variables often vary \citep{anselin2013spatial}. This phenomenon contradicts the assumptions of ordinary kriging, highlighting the need for interpolation methods specifically designed for non-stationary features. 

Many methods have been developed to enhance the Kriging model to account for second-order spatial non-stationarity and improve interpolation accuracy. The first and most straightforward solution for handling spatial second-order non-stationarity is to divide the non-stationary surface into several homogeneous subregions \citep{luo2023generalized}. Within each subregion, spatial stationarity holds, allowing for separate application of Ordinary Kriging. A representative method following this approach is Stratified Kriging (StK). However, modeling at partition boundaries can introduce discontinuities in the prediction results. 
To do the stratified modelling while maintain the smooth of the prediction result over space, some methods have been proposed to comprehensively model spatial relationships, enabling the construction of semivariogram functions for different strata. Notable examples include P-MSN \citep{gao2020spatial} and the Generalized Heterogeneous Model (GHM) \citep{luo2023generalized}. P-MSN and GHM provide effective stratified modeling solutions with improved spatial continuity and accuracy compared to ordinary StK. However, they still suffer from the inherent drawbacks of stratified modeling, namely, the effectiveness of this approach heavily depends on the reliability of spatial partitioning algorithms. Furthermore, partition-based modeling reduces the number of available samples within each subregion, which makes it challenging to fit reliable semivariogram functions \citep{liu2021geographical}.

The second solution is to model spatial non-stationary patterns by decomposing the predicted values of geographic variables into two components: a trend surface and spatial variability \citep{hengl2007regression}. The trend surface is assumed to be fitted by a deterministic function, while the spatial variability represents a random process that geostatistical methods can predict. Different methods are developed to model the trend \citep{hengl2007regression}. When non-stationarity varies by region, location information can be used to model the spatial trend, leading to the development of Universal Kriging \citep{stein1991universal}. In this approach, a deterministic polynomial function is used to represent the trend surface, while the residual spatial variability is captured through a stochastic component. Regression Kriging (RK) generalizes this idea by explicitly modeling the trend using a regression model, which can include multiple explanatory variables \citep{hengl2007regression}. RK first fits a regression model to predict the trend and then applies Kriging to the residuals \citep{hengl2004generic}. Because it integrates diverse information sources and allows for a flexible choice of trend models, regression kriging often achieves the highest predictive accuracy among Kriging-based interpolation methods.

However, incorporating explanatory variables significantly increases the cost of spatial interpolation and introduces more uncertainty. Additionally, for many geostatistical tasks, obtaining high quality explanatory variables remains a challenge \citep{yao2023extracting}. For example, a classic kriging interpolation task involves predicting mineral distribution. Despite the availability of numerous satellite sensors for obtaining remote sensing data, collecting explanatory variables for subsurface prediction is difficult. This presents a challenge for current spatial interpolation methods, including geostatistical models.

In this study, we are committed to developing a regression kriging method that do not require explanatory variables and can address spatial non-stationarity. In cases where explanatory variables cannot be obtained, we can utilize the spatial patterns of geographic variables to predict the trend surface. Specifically, we propose a feature free regression kriging model. For each observation point, a series of features can be extracted from the values and distance relationships of the remaining points in its neighboring region. These features are then used to construct a regression model for fitting the trend surface. Finally, the residuals are employed to construct Ordinary Kriging for spatial interpolation. To validate the performance of FFRK, we conducted a case study in a selected region of Australia, focusing on the prediction of three heavy metal concentrations. Furthermore, FFRK was compared against 17 classical spatial interpolation models.

\section{Feature-free regression kriging}

\subsection{Basic of Regression Kriging}

Regression Kriging is a hybrid spatial prediction method that combines a deterministic regression model with a geostatistical interpolation model. The fundamental idea of RK is to separately model the trend component and spatially correlated residuals, allowing for a more flexible and accurate spatial prediction.

\subsubsection{General Formulation of Regression Kriging}

Given an observed spatial variable \( Z(x) \) at location \( x \), RK decomposes the spatial variation into two components:

\begin{equation}
Z(x) = m(x) + \varepsilon(x)
\end{equation}
where \( m(x) \) represents the deterministic trend component, which captures large-scale variations and can be modeled using regression techniques,  \( \varepsilon(x) \) represents the spatially correlated residuals, which account for small-scale variations and are modeled using geostatistical interpolation.

\subsubsection{Trend Modeling via Regression}

The trend component \( m(x) \) is typically modeled using a regression function:

\begin{equation}
m(x) = g(\mathbf{X}(x))
\end{equation}
where \( g(\cdot) \) is a regression model, which can be a linear regression, random forest, or any machine learning algorithm. \( \mathbf{X}(x) \) is a vector of explanatory variables at location \( x \), which can include environmental, demographic, or geospatial factors.

The regression model is trained using known data points \( \{x_i, Z(x_i)\}_{i=1}^{n} \). Once trained, this regression model can predict the trend component at any location \( x_p \):

\begin{equation}
m(x_p) = g(\mathbf{X}(x_p))
\end{equation}
where $m(x_p)$ is the predicted trend at the location \( x_p \).

\subsubsection{Residual Modeling via Kriging}

After estimating the trend component, the residuals at known sample locations \( x_i \) are computed as:

\begin{equation}
\varepsilon(x_i) = Z(x_i) - m(x_i)
\end{equation}
where the $\varepsilon(x_i)$ is the residuals at the location \( x_i \).

These residuals are spatially correlated and modeled using geostatistical techniques. Using the fitted semivariogram  \( \gamma(h) \), residuals at unknown locations \( x_p \) are interpolated using ordinary kriging:

\begin{equation}
\varepsilon^*(x_p) = \sum_{i=1}^{n} w_i \varepsilon(x_i)
\end{equation}
where \( w_i \) are kriging weights derived from the semivariogram model.

\subsubsection{RK Prediction}

The predicted value at an unknown location \( x_p \) is obtained by combining the regression-predicted trend and the kriging-interpolated residual:

\begin{equation}
Z^*(x_p) = m(x_p) + \varepsilon^*(x_p)
\end{equation}
where $Z^*(x_p)$ is the predicted value at location \( x_p \), and

\subsection{Concept of FFRK}


The most important feature of RK is its ability to fully utilize explanatory variables to fit the trend, thereby improving the accuracy of ordinary kriging. However, its limitation is also evident: if explanatory variables are difficult to obtain, the method becomes unusable. 

\begin{figure*}[htbp]
    \centering
    \includegraphics[width=1\linewidth]{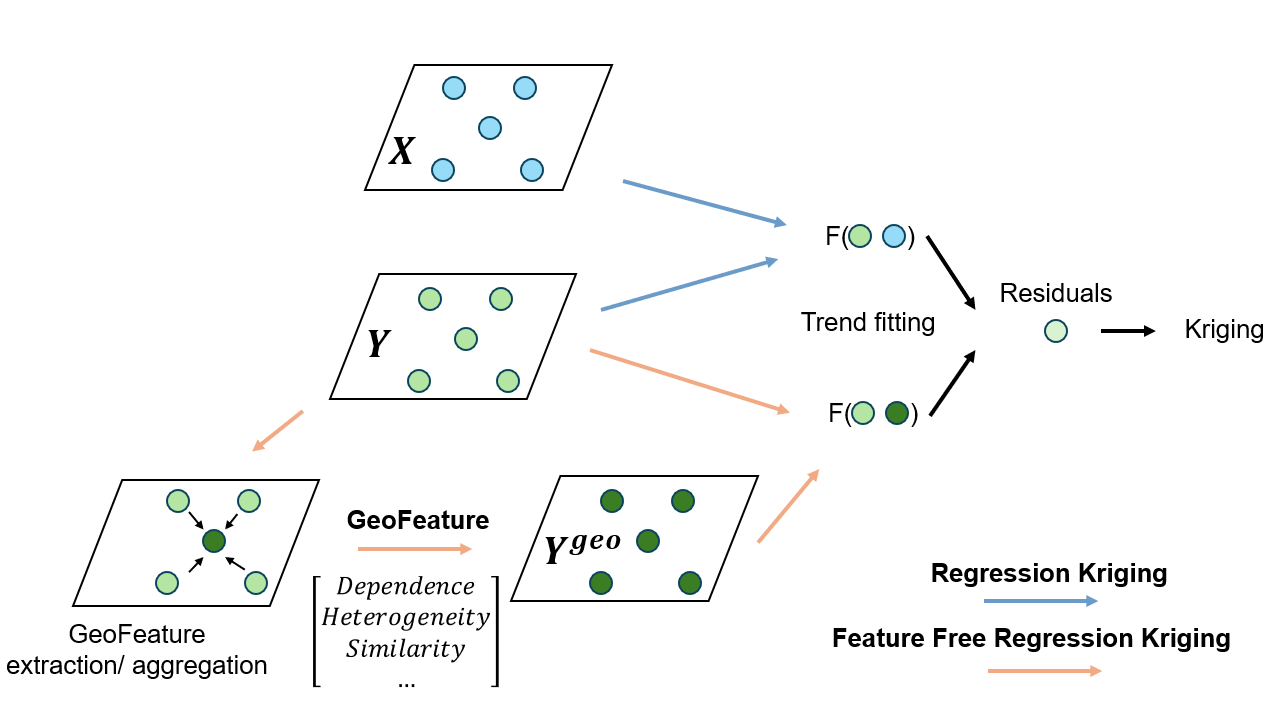}
    \caption{Schematic overview of the developed feature-free regression kriging (FFRK) for spatial interpolation, and its different wit the Regression Kriging}
    \label{fig:framework}
\end{figure*}

As is shown in Figure \ref{fig:framework}, the idea behind our proposed FFRK is straightforward: in the absence of explanatory variables, we replace them with the spatial distribution characteristics of the predicted variable itself to perform regression kriging. Therefore, FFRK follows the same algorithmic foundation as RK, with the only difference being in trend fitting—rather than constructing a regression model using explanatory variables, it is built based on geospatial features.

\begin{equation}
Z(x) = m(x) + \varepsilon(x)
\end{equation}

\begin{equation}
m(x) = g(\mathbf{F}(x))
\end{equation}
where: \( g(\cdot) \) is a predictive model, \( \mathbf{F}(x) \) is the extracted geo-feature vector.

Subsequently, the process of FFRK is entirely consistent with that of RK: the trend surface $m(x)$ is fitted based on geospatial features, followed by residual computation and kriging interpolation of the residuals.

\subsection{Geofeature extraction}

In spatial analysis, the interpolation accuracy highly depends on the feature representations of the data. the framework of FFRK can be combined with any kinds of geofeature, according to the specific application.

\begin{figure*}[htbp]
    \centering
    \includegraphics[width=1\linewidth]{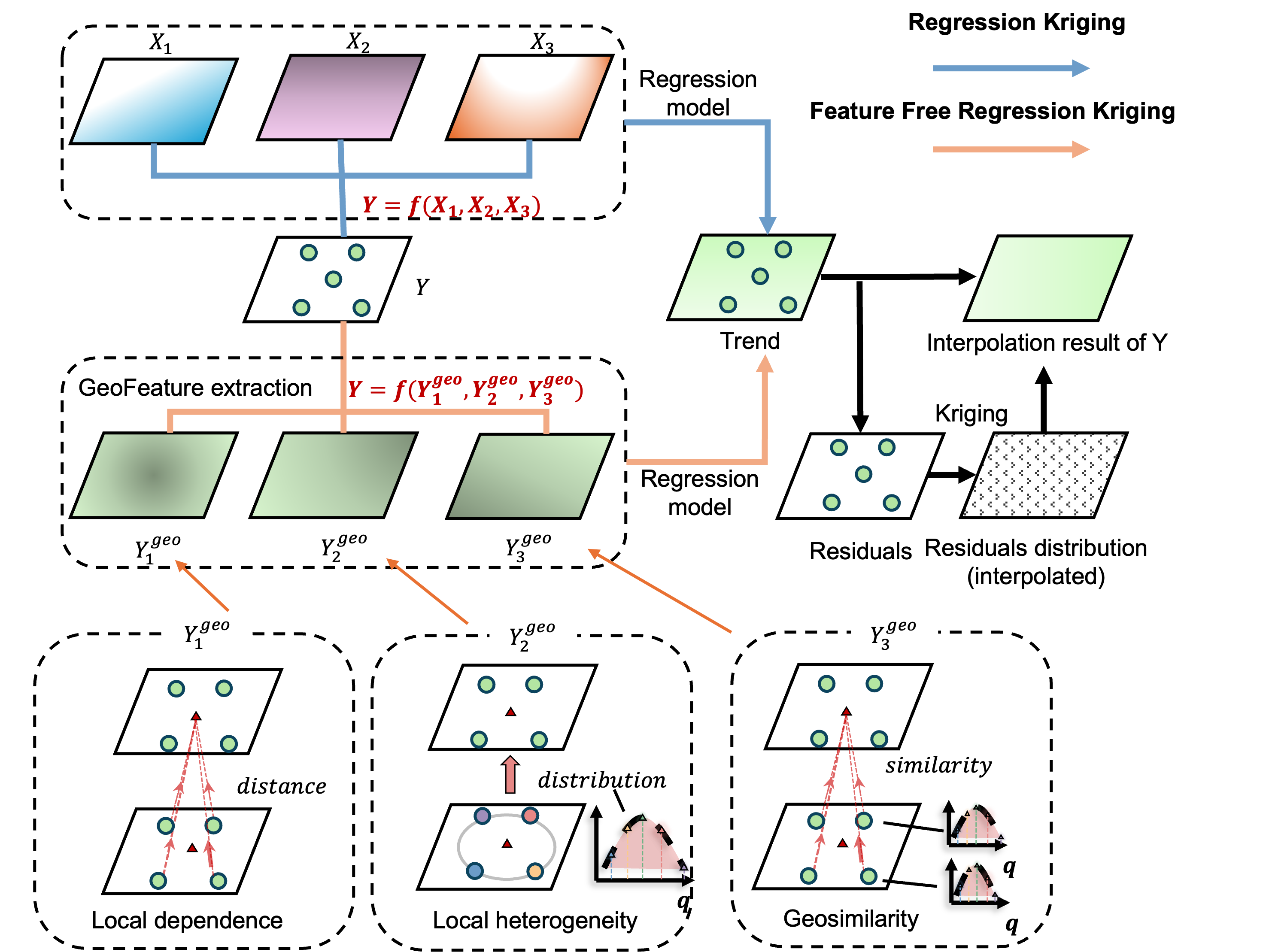}
    \caption{Three types of geofeatures for FFRK-based spatial interpolation used in this study, including local trend, local heterogeneity, and geosimilarity}
    \label{fig:geofeature}
\end{figure*}

In this study, as a showcase of the FFRK framework, we consider three distinct types of geospatial features to capture both local pattern and global pattern in spatial distributions (Figure \ref{fig:geofeature}). These features are designed to provide a more comprehensive representation of spatial processes and to enhance interpolation performance. The first is the local trend feature, extracted using IDW, which describes the general directional pattern of spatial structure in the neighborhood of each sample. This feature reflects the core principle of Tobler’s First Law of Geography—spatial proximity—and helps characterize the local continuity of spatial variables. The second is the local heterogeneity feature, which captures intra-regional variability and unevenness by computing statistical descriptors (e.g., quantiles) of neighboring observations. This feature highlights the spatial heterogeneity inherent in many geospatial phenomena. The third is the geosimilarity feature, which measures statistical similarity between sample points based on the local distributions used in the second feature. Unlike traditional distance-based methods, this similarity reflects environmental resemblance between distant locations, allowing the model to account for non-local spatial dependencies and enhancing its ability to identify globally consistent patterns.

Each prediction location \( x_p \) is associated with a feature vector:
\begin{equation}
\mathbf{F}(x_p) = \left[ \mathbf{F}_{\text{IDW}}(x_p), \mathbf{F}_{\text{SVD}}(x_p), \mathbf{F}_{\text{GOS}}(x_p) \right]
\end{equation}

where the three feature sets are computed as follows.

\subsubsection{Local Dependence: IDW Feature Extraction}
The IDW method estimates the value of a target point \( x_p \) based on the weighted average of surrounding observations \( x_i \), where the weight is inversely proportional to the distance:

\begin{equation}
Z^*(x_p) = \frac{\sum_{i=1}^{k} w_i Z(x_i)}{\sum_{i=1}^{k} w_i}
\end{equation}
where the weight \( w_i \) is defined as:
\begin{equation}
w_i = \frac{1}{d(x_p, x_i)^p}
\end{equation}
where \( d(x_p, x_i) \) is the Euclidean distance between the target point \( x_p \) and an observation \( x_i \).

IDW assumes a spatially homogeneous distribution without explicit trend modeling. However, it may introduce bias when a strong spatial trend exists. The extracted IDW-based feature for each point is:

\begin{equation}
\mathbf{F}_{\text{IDW}}(x_p) = [ f_{\text{IDW}}(x_p) ]
\end{equation}

\subsubsection{Local Heterogeneity: Spatially varying distribution}

We refer to the quantile-based characterization of local neighborhoods as the Spatially Varying Distribution (SVD) feature, which captures the local statistical structure of spatial data. The extracted feature vector \( \mathbf{F}_{\text{SVD}}(x_p) \) provides a structured representation of local heterogeneity. By incorporating SVD into spatial modeling, we can effectively account for local non-stationarity and spatial variability, making interpolation more adaptive to complex spatial structures.

First, given a target location \( x_p \), we define its neighborhood \( \mathcal{N}_k(x_p) \) as the set of its \( k \) nearest observed points:

\begin{equation}
\mathcal{N}_k(x_p) = \{ x_i \mid d(x_i, x_p) \text{ is among the } k \text{ smallest distances} \}
\end{equation}
where the distance \( d(x_i, x_p) \) is computed using the Euclidean metric.

Second, for each point \( x_p \), we compute the empirical quantiles of the variable \( Z(x) \) within its neighborhood \( \mathcal{N}_k(x_p) \):

\begin{equation}
Q_q(x_p) = \text{Quantile} \left( \{ Z(x_i) \mid x_i \in \mathcal{N}_k(x_p) \}, q \right)
\end{equation}
where \( Q_q(x_p) \) represents the \( q \)-th quantile of the observed values in the neighborhood.

Third, the final SVD feature vector for location \( x_p \) consists of multiple quantile values across different probability levels:

\begin{equation}
\mathbf{F}_{\text{SVD}}(x_p) = \left[ Q_{q_1}(x_p), Q_{q_2}(x_p), ..., Q_{q_d}(x_p) \right]
\end{equation}
where \( \{q_1, q_2, ..., q_d\} \) represents the predefined quantile levels (e.g., \( q \in \{0, 0.05, 0.10, ..., 1.0\} \)).

\subsubsection{Geosimilarity: GOS Feature Extraction}

The Geographically Optimal Similarity (GOS) method is selected to being as the third geofeature to estimate the target variable at unknown locations \citep{song2023geographically}. GOS identifies spatial configurations with similar structures to make predictions.


First, we define the spatial configuration of each observed location based on extracted features. For each observed location \( x_i \), we define its spatial configuration based on a set of extracted features $\mathbf{F}_{\text{SVD}}(x_p)$.

Second, the similarity between an unknown location \( x_p \) and an observed location \( x_i \) is defined as:

\begin{equation}
S(x_i, x_p) = P \left\{ E_j ( f_{\text{SVD},j}(x_i), f_{\text{SVD},j}(x_p) ) \right\}
\end{equation}
where \( E_j \) is the similarity function between the \( j \)-th feature of the observed and unknown locations, \( P \) is an aggregation function to determine overall similarity.

For continuous spatial features, we define \( E_j \) as:

\begin{equation}
E_j(x_i, x_p) = \exp \left( - \frac{( F_{\text{SVD},j}(x_i) - F_{\text{SVD},j}(x_p) )^2}{2\sigma_j^2} \right)
\end{equation}
where \( \sigma_j \) represents the standard deviation of feature \( j \).

Third, instead of using all observations for prediction, GOS selects only the most similar locations. The optimal similarity threshold \( S_\lambda \) is determined by minimizing prediction errors:

\begin{equation}
\lambda = \arg\min_{\kappa} \text{RMSE}(\kappa)
\end{equation}
where \( \kappa \) is the proportion of the most similar samples used for prediction, and RMSE is the root mean square error from cross-validation.

Fourth, the prediction at \( x_p \) is computed as:

\begin{equation}
F_{\text{GOS}}(x_p)= \frac{\sum_{i \in \mathcal{N}_\lambda} S_\lambda(x_i, x_p) Z(x_i)}{\sum_{i \in \mathcal{N}_\lambda} S_\lambda(x_i, x_p)}
\end{equation}
where \( \mathcal{N}_\lambda \) is the set of selected observations with similarity above \( S_\lambda \).

\subsection{Regression kriging with geofeatures}

Combining all three components, we define the final spatial feature vector:

\begin{equation}
\mathbf{F}(x_p) = \left[ F_{\text{IDW}}(x_p),  F_{\text{SVD}}(x_p), F_{\text{GOS}}(x_p) \right]
\end{equation}
where:
 \( F_{\text{IDW}}(x_p) \) captures local dependence.
 \( \mathbf{F}_{\text{SVD}}(x_p) \) captures local heterogeneity using statistical quantiles.
 \( F_{\text{GOS}}(x_p) \) captures global similarity based on regression over spatially similar regions.

The final feature vector has a total of:

\begin{equation}
\text{dim}(\mathbf{F}) = 1 + d_{\text{SVD}} + 1
\end{equation}
where:
 \( 1 \) represents the IDW feature.
 \( d_{\text{SVD}} \) represents the number of quantile-based SVD features (which can be adjusted based on resolution).
 \( 1 \) represents the GOS similarity feature.

A machine learning regression model \( g(\mathbf{F}) \) is trained on known observations, which captures the large-scale trend:

\begin{equation}
Z_{\text{trend}}(x) = g(\mathbf{F}(x))
\end{equation}
where $Z_{\text{trend}}(x)$ is the trend value at the location x.


The final step is to interpolate the residuals and combine them with the predicted trend using ordinary kriging.


\section{Case study: mapping trace elements with FFRK}

\subsection{Study area and data}

\begin{figure*}[htbp]
    \centering
    \includegraphics[width=1\linewidth]{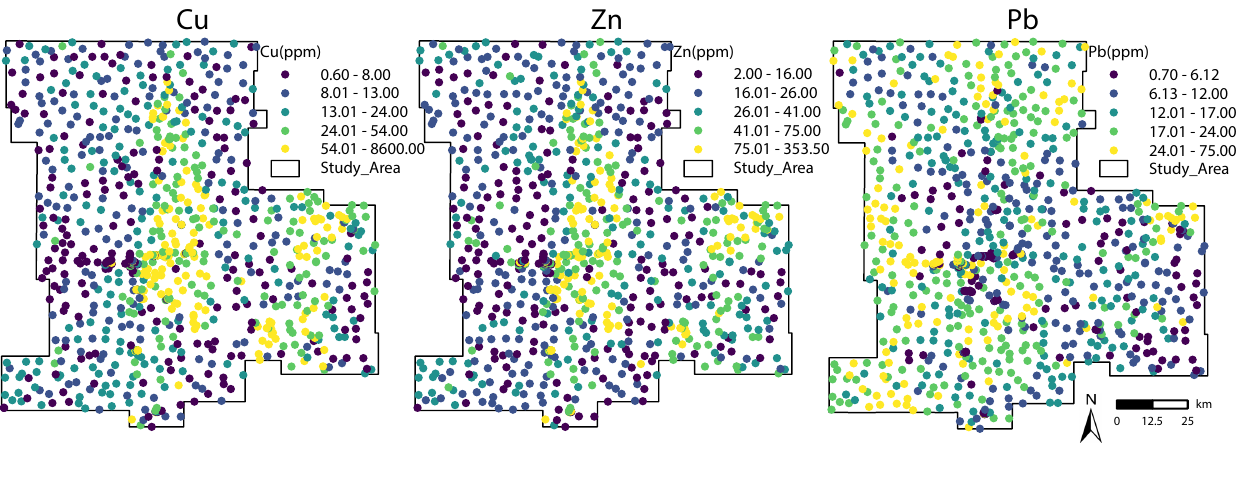}
    \caption{Spatial distributions of trace elements: Cu, Zn, and Pb}
    \label{fig:var}
\end{figure*}

In this work, we use trace element data of Cu, Pb, and Zn, from one region of Australia to test the performance of FFRK.  We concentrated on the geographic variability of the three elements for the trace element distribution because these elements are well known to be important markers of environmental contamination and are essential for evaluating ecological health. The spatial distributions of trace elements are shown in Figure \ref{fig:var}.

We included nine environmental explanatory variables in alphabetical order to examine the factors influencing their spatial patterns: slope, road network density (Road), normalized difference vegetation index (NDVI), distance to major roads (MainRd), distance to lithology (Dlith), distance to fault lines (Dfault), soil organic carbon (SOC), soil pH levels, and water distribution. These variables were selected for their relevance in characterizing and predicting the distribution of trace elements. The calculation of the variables, such as Dlith, Dfault, is referenced in  \citep{song2023geographically}. Table~\ref{tab:env_stats} summarizes the descriptive statistics of all explanatory variables, including the mean (Mean), minimum (Min), median, maximum (Max), standard deviation (SD), and coefficient of variation (CV).

\begin{table}[h]
    \centering
    \caption{Descriptive statistics of environmental variables}
    \resizebox{\textwidth}{!}{%
    \begin{tabular}{l l rrrrrr}
        \hline
        Variable & Code & Mean & Min & Median & Max & SD & CV \\
        \hline
        Distance to water(km) & Water & 1.114 & 0.000 & 0.648 & 7.696 & 1.188 & 1.067 \\
        Distance to mainroads(km) & MainRoads & 19.991 & 0.010 & 17.485 & 58.371 & 14.219 & 0.711 \\
        Distance to roads(km) & Roads & 9.817 & 0.002 & 7.953 & 50.141 & 8.363 & 0.852 \\
        Distance to mine(km) & MineKm & 14.785 & 0.025 & 11.425 & 56.638 & 12.117 & 0.820 \\
        Slope(degree) & Slope & 0.275 & 0.008 & 0.246 & 1.712 & 0.161 & 0.585 \\
        Normalized difference vegetation index  & NDVI & 0.178 & 0.062 & 0.180 & 0.251 & 0.024 & 0.135 \\
        Soil organic carbon & SOC & 0.868 & 0.686 & 0.870 & 1.066 & 0.053 & 0.061 \\
        Soil pH & pH & 5.741 & 5.113 & 5.753 & 6.178 & 0.178 & 0.031 \\
        Distance to lithology (km) for Cu & $Dlith_{Cu}$ & 9.132 & 0.000 & 7.147 & 39.839 & 7.769 & 0.851 \\
        Distance to lithology (km) for Zn & $Dlith_{Zn}$ & 8.150 & 0.000 & 6.229 & 39.839 & 7.637 & 0.937 \\
        Distance to lithology (km) for Pb & $Dlith_{Pb}$ & 3.363 & 0.000 & 2.627 & 15.946 & 3.020 & 0.898 \\
        Distance to fault(km) for Cu & $Dfault_{Cu}$ & 16.070 & 0.001 & 13.285 & 54.765 & 12.150 & 0.756 \\
        Distance to fault(km) for Pb & $Dfault_{Pb}$ & 12.017 & 0.003 & 11.111 & 43.471 & 8.090 & 0.673 \\
        Distance to fault(km) for Zn & $Dfault_{Zn}$ & 14.174 & 0.001 & 11.844 & 52.697 & 10.851 & 0.766 \\
        Elevation(m) & Elevation & 482.699 & 398.050 & 485.128 & 588.649 & 36.700 & 0.076 \\
        Aspect(degree) & Aspect & 171.645 & 0.751 & 174.126 & 358.712 & 90.576 & 0.528 \\
        \hline
    \end{tabular}%
    }
    \label{tab:env_stats}
\end{table}

\subsection{Experiment design}

\begin{figure*}[htbp]
    \centering
    \includegraphics[width=1\linewidth]{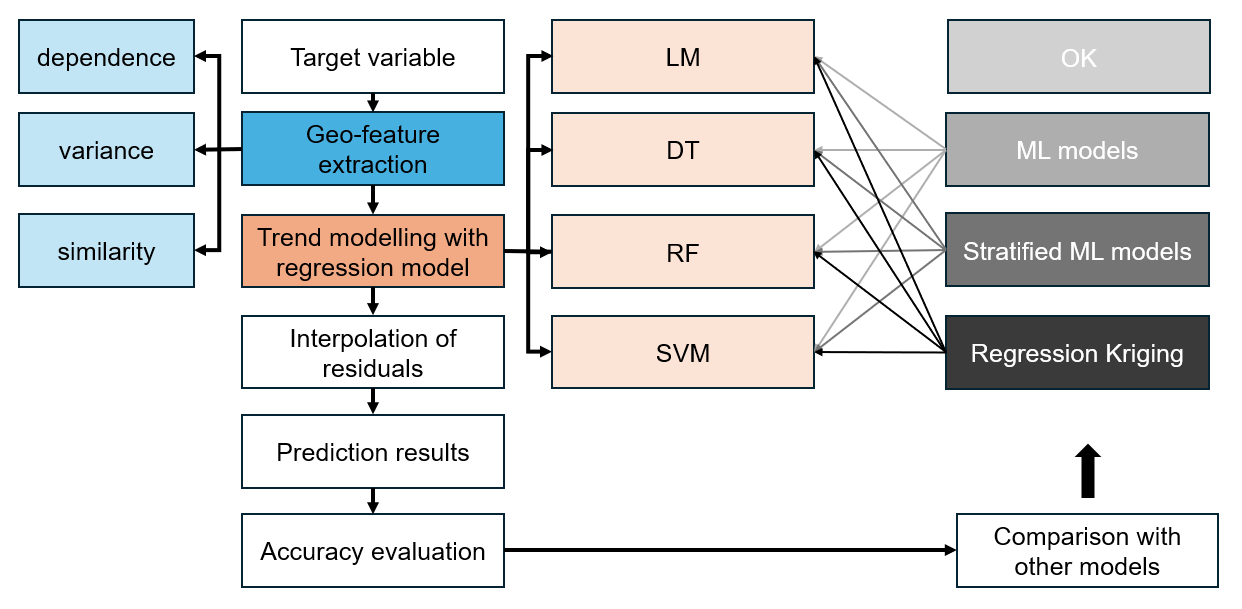}
    \caption{Experimental design for evaluating the performance of FFRK in comparison with other interpolation models}
    \label{fig:design}
\end{figure*}

Figure \ref{fig:design} illustrates the experimental design of this study, which aims to validate the reliability of the proposed FFRK method. For the target variable, we computed its three-dimensional geofeatures, including dependence, heterogeneity, and similarity. Subsequently, we selected four types of machine learning models—linear model (LM), decision tree (DT), random forest (RF), and support vector machine (SVM)—to predict the trend of the target variable. After obtaining the residuals by comparing the predicted trend with observations, we applied ordinary kriging for spatial interpolation of the residuals. Finally, the predicted trend and interpolated residuals were combined to produce the prediction. 

We compare our proposed method with several representative baselines, including classical geostatistical approaches, machine learning models, and their stratified and hybrid variants. First, Ordinary Kriging (OK) serves as the foundational geostatistical method and is used as a baseline for performance comparison. Second, we consider machine learning models, which directly establish predictive relationships between explanatory variables and the target variable. In this study, we employ four commonly used models: Linear Regression (LM), Decision Tree Regression (DT), Random Forest Regression (RF), and Support Vector Machine Regression (SVM).

Third, stratified models are introduced to better account for spatial heterogeneity. These models partition the study area into multiple subregions based on homogeneity analysis, with separate machine learning models trained within each region. The same four algorithms (LM, DT, RF, and SVM) are applied in each subregion. The spatial partitioning is conducted using a decision tree, which segments the domain based on geographic coordinates (longitude and latitude) as well as trace element values. In addition to stratified machine learning models, we also include stratified kriging, resulting in a total of five stratified models.

Fourth, we evaluate Regression Kriging (RK), a hybrid method that combines trend modeling and spatial interpolation. Specifically, a machine learning model (LM, DT, RF, or SVM) is used to estimate the trend, and the residuals are then interpolated using kriging. Finally, we present our proposed method, Feature-Filtered Regression Kriging (FFRK). Similar to RK, FFRK follows a two-step structure but incorporates a feature selection process prior to regression. The four variants of FFRK correspond to the four regression models used: FFRK (LM), FFRK (DT), FFRK (RF), and FFRK (SVM).

In total, we compared 18 models. Among them, only Ordinary Kriging (OK), Stratified Kriging (StK) and four FFRK models do not require any explanatory variables, whereas the other three model categories require nine explanatory variables as inputs.

In the FFRK model, two hyperparameters are involved. The first is K, which defines the number of neighboring sample points used to compute geo-features and fit the semivariogram. The second is the quantile interval, used in the second geo-feature to characterize the statistical information within the neighborhood of each sample point. In this study, we set K = 15 and the quantile interval = 5\%, following the settings adopted in a previous study \citep{song2023geographically}. A sensitivity analysis of these two hyperparameters and their impact on the performance of FFRK is presented in Section 3.4.2.

\subsection{Cross-validation and Model Evaluation}

In this study, we adopted 10-fold cross-validation to evaluate the prediction performance and generalization capability of the proposed FFRK method and baseline models. Specifically, the dataset was randomly divided into ten subsets of equal size. In each iteration, nine subsets served as the training set, while the remaining subset was reserved as the validation set. This training-validation procedure was repeated ten times, with the averaged results across the ten iterations representing the final model performance. This approach effectively mitigates the instability arising from a single random partition and enhances the robustness of model evaluation.

In particular, for Stratified Models, we employed a regression tree to partition the study area spatially, and subsequently performed stratified sampling based on the distribution of the target variable. This ensured consistent internal data distribution across each of the ten folds constructed for cross-validation.

Moreover, the cross-validation process was utilized for model parameter optimization. Optimal parameters for each machine learning model were determined by minimizing the mean RMSE obtained from cross-validation. Finally, each model was retrained once using the complete dataset to derive globally optimal parameters, and the resulting optimized models were preserved for subsequent global spatial predictions.



\subsection{Results}

\subsubsection{Accuracy evaluation}

\begin{figure*}[htbp]
    \centering
    \includegraphics[width=1\linewidth]{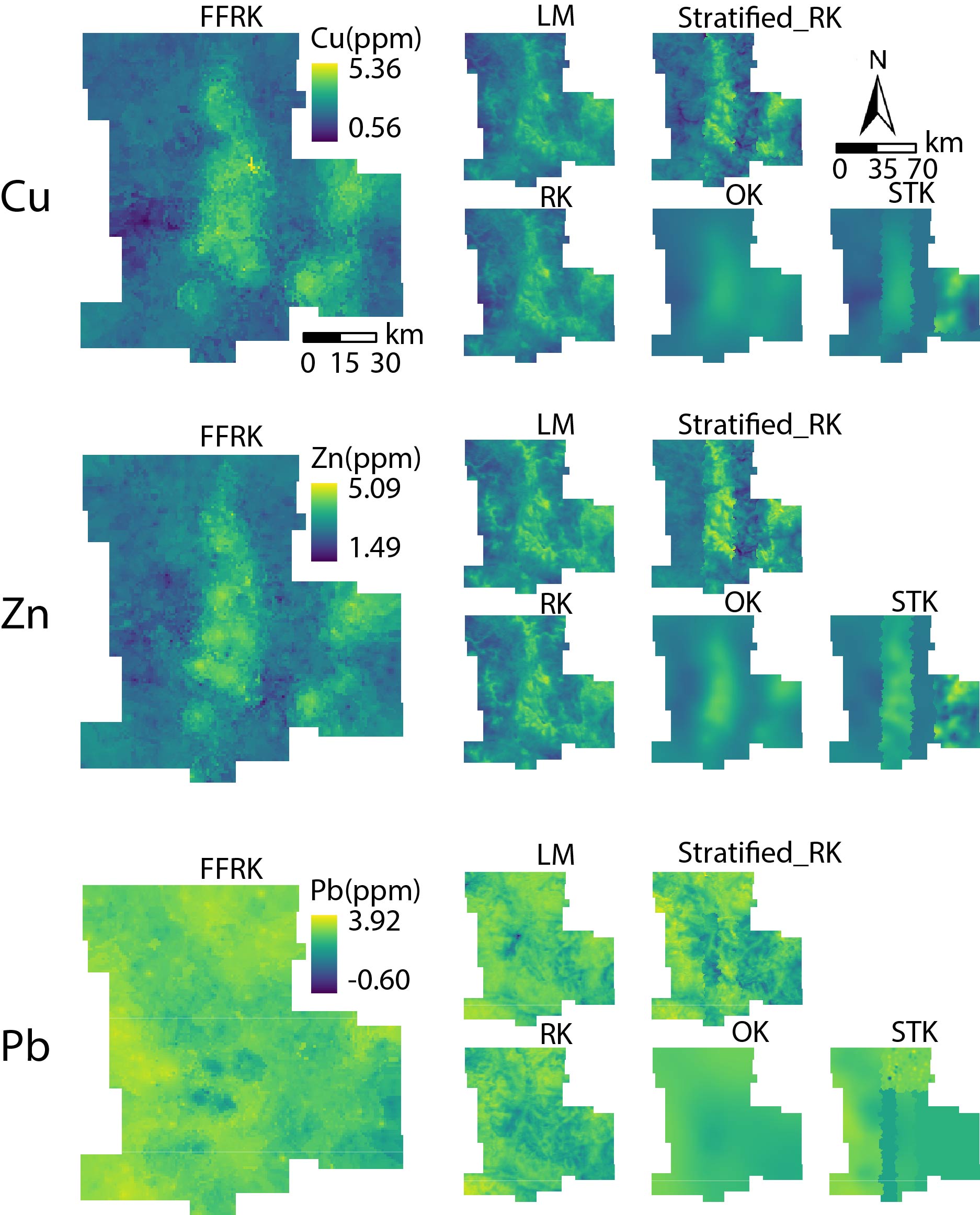}
    \caption{Spatial distributions of the predicted results for three trace elements—Cu, Zn, and Pb—are shown using six prediction models: FFRK, LM, RK, Stratified RK, OK, and STK}
    \label{fig:map}
\end{figure*}

Figure \ref{fig:map} presents the spatial interpolation results of FFRK alongside several other regression-based models. We selected Linear Model (LM) as the representative machine learning regression method, and also included Regression Kriging (RK) based on LM, as well as Stratified Kriging (StK). Additionally, we compared the results of Ordinary Kriging (OK) and StK. Our observations show that OK and StK produce the smoothest interpolation surfaces among the methods evaluated. However, this smoothness comes at the cost of lacking fine-grained spatial detail. In contrast, FFRK produces spatial patterns that closely resemble those of LM and RK, capturing more nuanced local variations.

\begin{figure*}[htbp]
    \centering
    \includegraphics[width=1\linewidth]{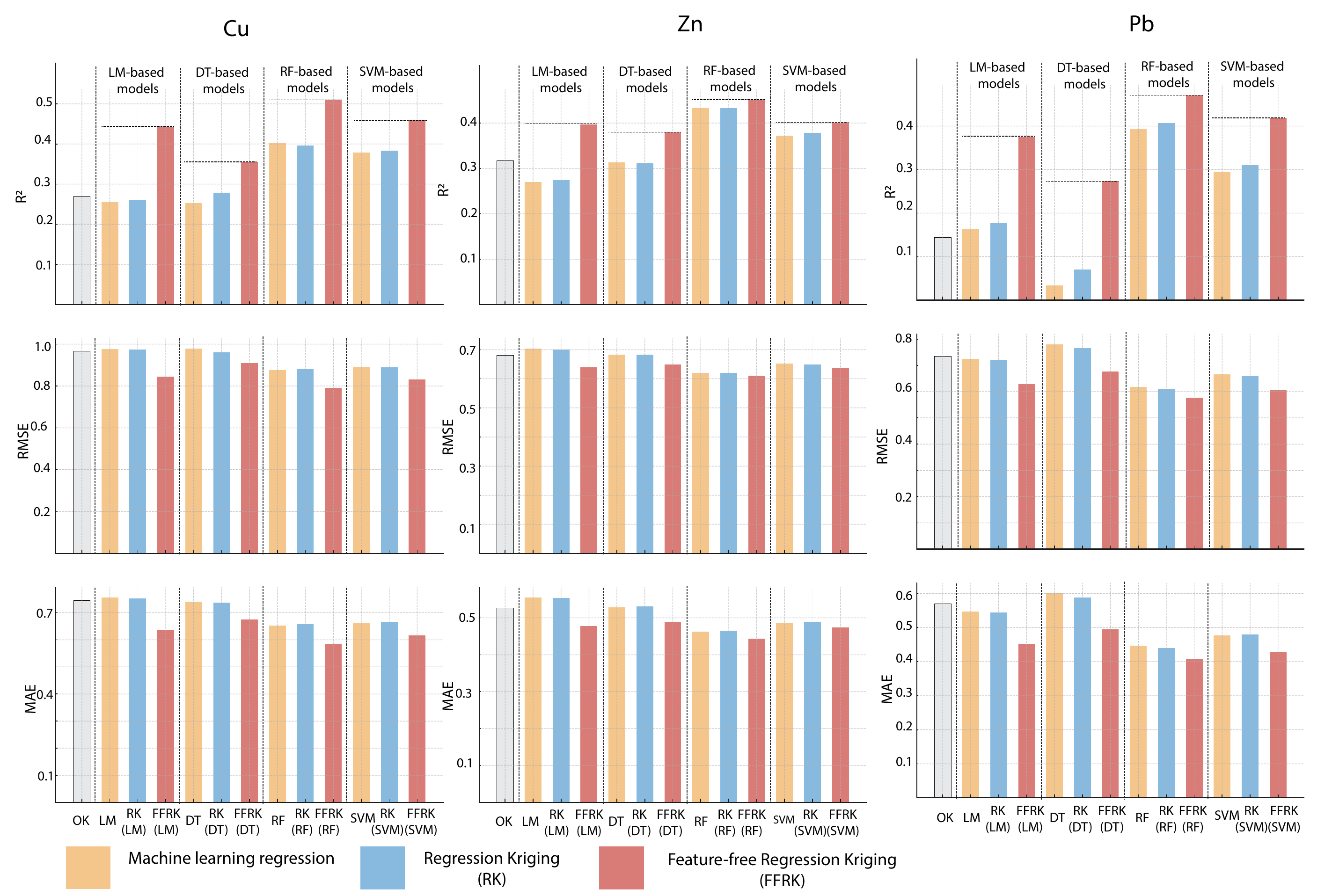}
    \caption{Evaluation results ($R^2$, $RMSE$, and $MAE$) for the 13 global models}
    \label{fig:evaluation}
\end{figure*}

Figure \ref{fig:evaluation} presents the accuracy results of 13 global modeling approaches for predicting the concentrations of three heavy metals, evaluated using R², RMSE, and MAE. Among these methods, 12 involve the use of machine learning (ML) models, with the exception of ordinary kriging (OK). To ensure a fair comparison, we categorized the methods into four groups based on the ML model used for trend modeling: LM-based, DT-based, RF-based, and SVM-based. 

The results indicate that the FFRK method, despite not incorporating any explanatory variables, achieves significantly higher accuracy than both standalone ML models and regression models based on ML modeling. The latter two approaches utilized nine explanatory variables and a large dataset but failed to yield higher prediction accuracy. For example, in Cu prediction, the R² values of FFRK with LM-, DT-, RF-, and SVM-based trend modeling exceeded those of regression kriging using the same ML models by 73.72\%, 40.31\%, 26.87\%, and 21.27\%, respectively. The most significant improvement was observed in Pb prediction, where FFRK with DT-based trend modeling achieved an R² of 0.27, surpassing regression kriging and the DT model by 0.07 and 0.03, respectively.

In Table \ref{tab:model_means}, we present the results of five stratified modeling approaches. First, we partitioned the spatial domain into multiple homogeneous subregions based on the values of explanatory variables. Within each subregion, we decomposed and fitted separate models to perform spatial predictions. Finally, we evaluated the overall prediction accuracy across the entire study area. To assess the impact of stratification, we experimented with different numbers of partitions, ranging from 2 to 10, and calculated the average accuracy. The results indicate that compared to global modeling approaches, including OK and various machine learning models, stratified modeling can effectively improve prediction accuracy, as discussed in the Introduction. However, its performance still falls short of that achieved by the FFRK method. In some cases, stratification resulted in over-segmentation, where certain subregions contained too few observations, leading to model underfitting and significantly reduced predictive performance. Additionally, stratified modeling introduced abrupt and unrealistic discontinuities in the final spatial interpolation results, which arose due to artificial boundaries between partitions.

\begin{table}[h]
    \centering
    \caption{Mean R², MAE, and RMSE for Different Models and Elements (K=2 to 10)}
    \begin{tabular}{llrrr}
        \hline
        Element &          Model &  Mean R² &  Mean MAE &  Mean RMSE \\
        \hline
        \multirow{5}{*}{Cu} & Stratified\_Kriging &   0.3584 &    0.6720 &     0.8964 \\
                            & Stratified\_LM &   0.3738 &    0.6699 &     0.8978 \\
                            & Stratified\_DT &   0.3009 &    0.7114 &     0.9486 \\
                            & Stratified\_RF &   0.4115 &    0.6406 &     0.8705 \\
                            & Stratified\_SVM &   0.4205 &    0.6318 &     0.8637 \\
        \hline
        \multirow{5}{*}{Zn} & Stratified\_Kriging &   0.3791 &    0.4830 &     0.6287 \\
                            & Stratified\_LM &  -6.4581 &    0.5559 &     1.5290 \\
                            & Stratified\_DT &   0.3546 &    0.5054 &     0.6613 \\
                            & Stratified\_RF &   0.4367 &    0.4575 &     0.6179 \\
                            & Stratified\_SVM &   0.4147 &    0.4725 &     0.6298 \\
        \hline
        \multirow{5}{*}{Pb} & Stratified\_Kriging &   0.2747 &    0.5277 &     0.6958 \\
                            & Stratified\_LM &   0.1043 &    0.5040 &     0.7462 \\
                            & Stratified\_DT &   0.2178 &    0.5296 &     0.7041 \\
                            & Stratified\_RF &   0.4063 &    0.4399 &     0.6137 \\
                            & Stratified\_SVM &   0.3776 &    0.4474 &     0.6283 \\
        \hline
    \end{tabular}
    \label{tab:model_means}
\end{table}

The scatter plots comparing the observed and predicted values for all models are presented in the appendix (Figures \ref{fig:cu}, \ref{fig:zn}, and \ref{fig:pb}). It can be seen that the scatter plots of the FFRK models, which exhibit the highest accuracy, are the closest to the 1:1 line.

\subsubsection{Sensitivity analysis for the hyperparameter in FFRK}

\begin{figure*}[htbp]
    \centering
    \includegraphics[width=1\linewidth]{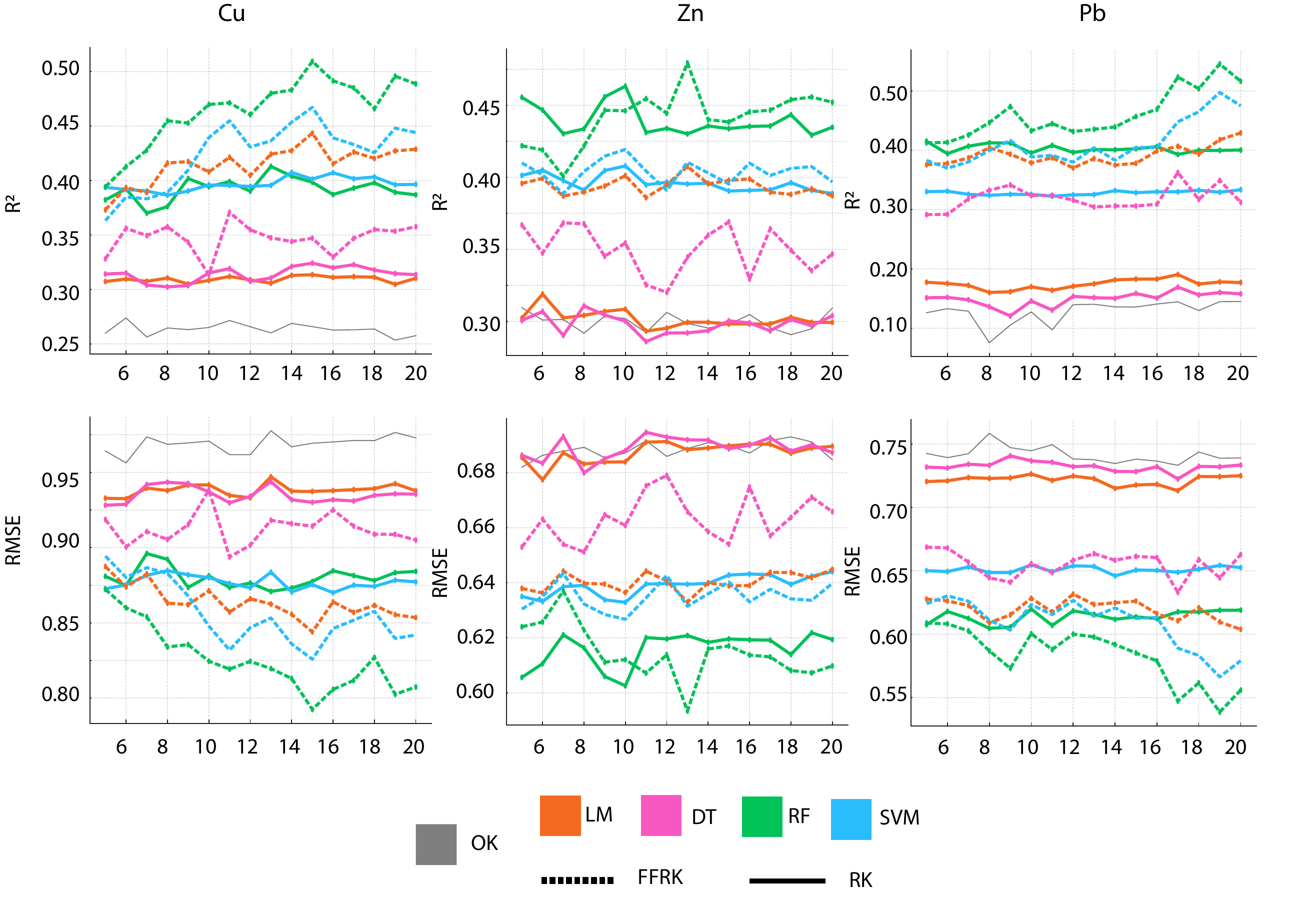}
    \caption{The sensitivity analysis for the K value}
    \label{fig:kvalue}
\end{figure*}

In Kriging-based methods, an hyperparameter is the number of neighboring points used when fitting the semivariogram and performing interpolation. In the previous experiments, to ensure a fair comparison among different methods, we set the number of nearest neighbors (k) to 15 for all models. In this section, we assess the sensitivity of different methods to k. Figure \ref{fig:kvalue} presents the accuracy of FFRK and RK models under varying numbers of neighboring points. 

The results indicate that for the same type of ML model used in trend modeling, FFRK (dashed line) consistently outperforms RK (solid line) across most k values, demonstrating higher R² and lower RMSE. However, there are a few exceptions. For instance, when k is less than 11 in Zn prediction, the RF-based RK model achieves higher accuracy than the RF-based FFRK model. Despite this, FFRK based on LM and DT still significantly outperforms RK when k is below 11. This discrepancy may be attributed to the distribution characteristics of Zn: when k is small, the geofeatures used by FFRK may not provide sufficient information to properly fit the RF model. In contrast, the RK method, which utilizes nine explanatory variables, has enough information for RF to achieve better fitting. However, for the relatively simpler LM and DT models, the geofeatures in FFRK provide sufficient information, enabling FFRK to surpass RK in predictive performance.

\begin{figure*}[htbp]
    \centering
    \includegraphics[width=1\linewidth]{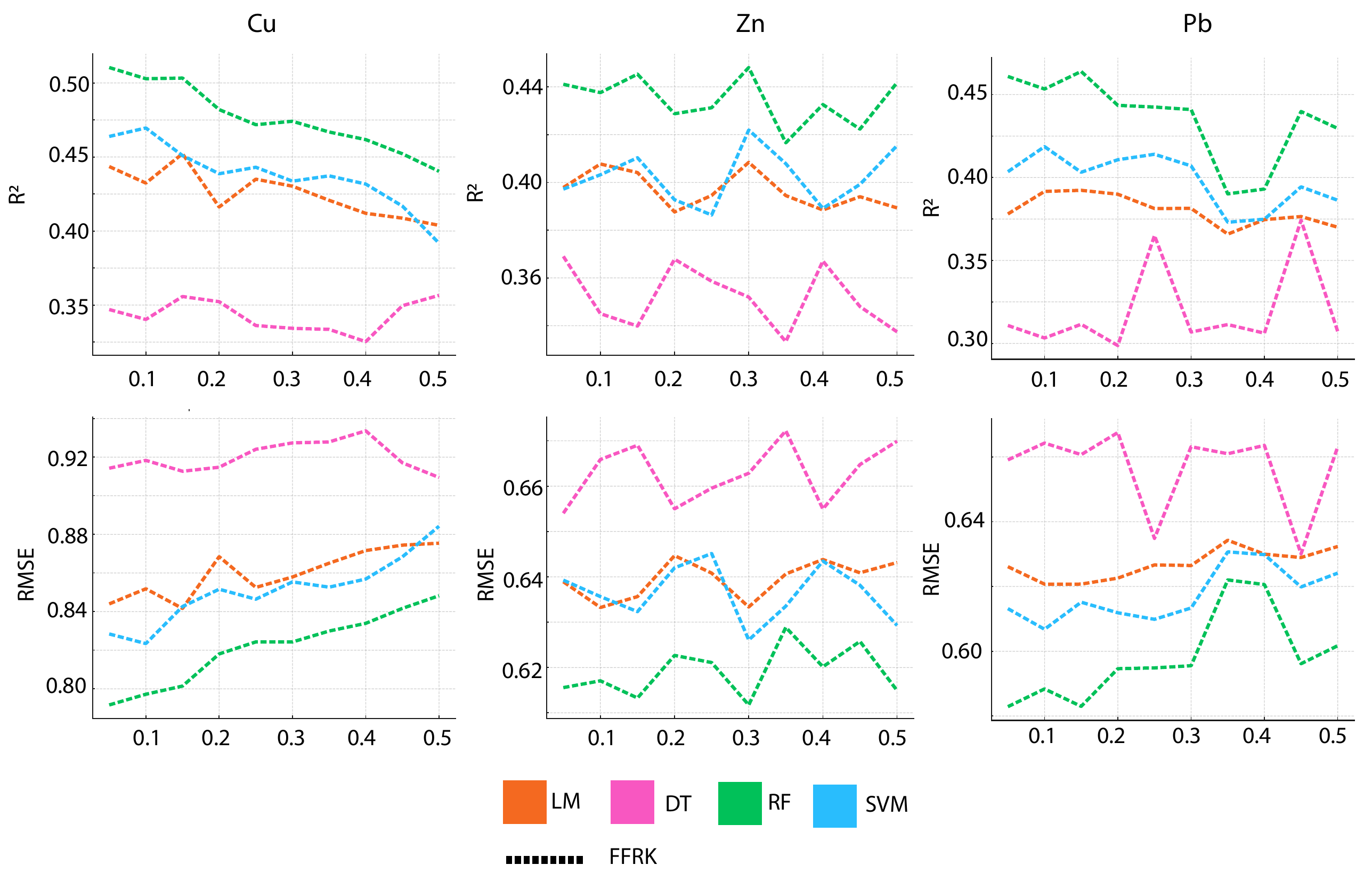}
    \caption{The sensitivity analysis for the quantile step}
    \label{fig:qvalue}
\end{figure*}

In the second category of geofeatures, we sample the value distribution of surrounding sample points for each predicted location. This sampling is based on fixed quantile intervals, such as 1\% or 5\%. In the previous experiments, we selected a 5\% quantile interval, resulting in 20 features in this geofeature category. To assess the sensitivity of FFRK to this parameter, we varied the quantile interval from 5\% to 50\% and evaluated its impact on the results. Figure \ref{fig:qvalue} shows that the FFRK method is relatively robust to changes in the quantile sampling interval. The maximum fluctuation in R² is approximately 0.05, while the highest variation in RMSE is only around 0.02.

\subsubsection{Generalized regression kriging}

In previous experiments, we have demonstrated that FFRK, despite not requiring any explanatory variables, can outperform RF models that utilize multiple explanatory variables solely through the extraction of geofeatures. Here, we propose a new question: if geofeatures and explanatory variables are integrated together for trend modeling, can the predictive performance be further improved? We define this approach, which incorporates both geofeatures and explanatory variables, as generalized regression kriging (GRK).  

\begin{figure*}[htbp]
    \centering
    \includegraphics[width=1\linewidth]{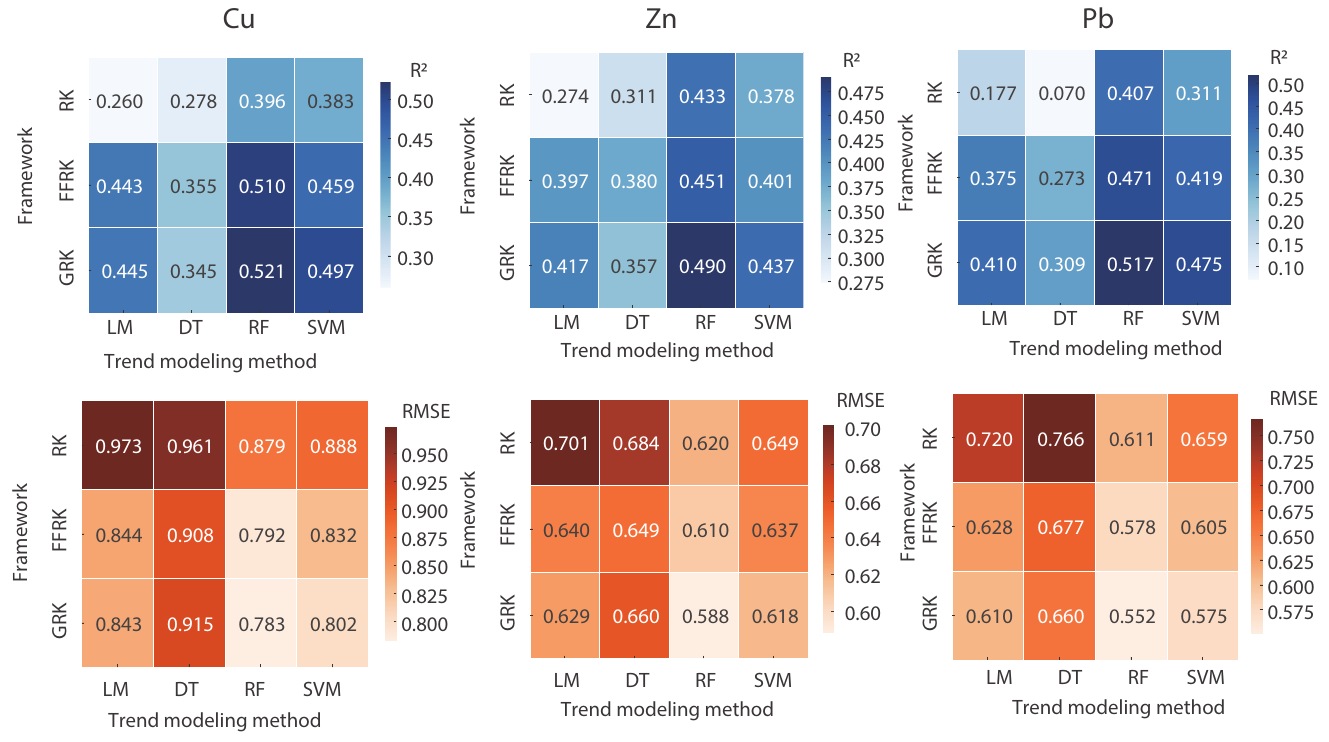}
    \caption{Accuracy of three framework: regression kriging, FFRK, and Generalized Regression Kriging (GRK)}
    \label{fig:grk}
\end{figure*}

The results, shown in Figure \ref{fig:grk}, indicate that across three different datasets and 12 prediction tasks (based on four ML models for trend modeling), GRK outperforms FFRK in 10 cases and significantly surpasses RK in all cases. The two exceptions occur in Cu and Zn prediction tasks when DT is used for trend modeling, where FFRK, relying solely on geofeatures, achieves a slightly higher R² than GRK, which includes explanatory variables. 

These findings suggest that incorporating more information into trend modeling enhances the performance of regression kriging models. The proposed GRK model is particularly suitable for spatial interpolation tasks where explanatory variables are available. Compared to conventional regression kriging, which relies solely on explanatory variables, integrating geofeatures into trend modeling leads to substantial improvements in predictive accuracy.

\section{Discussion}

Regression Kriging is a highly effective spatial interpolation model for addressing second-order spatial non-stationarity, which fits a trend surface based on explanatory variables and then performs kriging on the residuals. However, in many spatial interpolation tasks, it is often difficult to obtain explanatory variables, or to find variables that are sufficiently suitable, which greatly limits the applicability of regression kriging.

This work proposes a novel interpolation method for regression kriging under spatial non-stationarity without relying on explanatory variables, called FFRK. FFRK substitutes the traditional need for explanatory variables in trend surface fitting by creating features—termed geofeatures—based on the spatial distribution patterns of geographic variables. In this study, we introduce three categories of geofeatures, derived respectively from spatial dependence, spatial heterogeneity, and geographic similarity. We apply FFRK to a case study involving the spatial distribution of three heavy metal concentrations in a region of Australia. We also compare the performance of FFRK against 17 other interpolation methods of various types, demonstrating that FFRK achieves the highest predictive accuracy.

FFRK achieves superior performance without using any explanatory variables, outperforming other methods that do use explanatory variables, including machine learning models and regression kriging. This showcases the advantage of using features derived from the spatial distribution itself in regression kriging. However, the advantage does not imply that such spatial distribution features can entirely replace information from traditional explanatory variables. For the specific case study in this paper—predicting heavy metal concentrations—it is particularly difficult to identify suitable, large-scale, readily available explanatory variables. In most spatial prediction tasks, large-scale remote sensing observations are typically relied upon. However, subsurface distributions are often poorly represented by surface-level features detectable through remote sensing. This is precisely the application scenario that FFRK aims to address. As shown in Figure \ref{fig:grk}, when both explanatory variables and geofeatures are used together (a model we refer to as GRK), the prediction accuracy generally improves further compared to FFRK alone.

\section{Conclusion}

This study proposes FFRK, a regression kriging approach that eliminates the need for explanatory variables by extracting geospatial features from the response variable itself. Addressing key limitations in existing methods—such as reliance on external variables and assumptions of spatial stationarity—FFRK captures spatial dependence, local heterogeneity, and geographic similarity to construct a robust trend surface.
We demonstrate the effectiveness of FFRK through a spatial prediction task involving the distribution of three heavy metal concentrations in a region of Australia. FFRK outperforms traditional kriging, machine learning, and stratified methods, offering a practical solution for large-scale, non-stationary spatial interpolation tasks where explanatory data are scarce or unavailable.
This work contributes to GIScience by demonstrating that spatial features alone can effectively support predictive modeling. However, FFRK’s performance depends on the spatial density of observations. Future work will explore its extension to spatiotemporal settings and automated feature learning for broader applicability.

\section*{Disclosure Statement}

No conflict of interest exists in this manuscript, and the manuscript was approved by all authors for publication.

\baselineskip12pt
\bibliographystyle{elsarticle-harv} 
\bibliography{02_references.bib}


\section*{Appendix}
\appendix
\renewcommand{\thefigure}{S\arabic{figure}}  
\setcounter{figure}{0}   

\begin{figure*}[htbp]
    \centering
    \includegraphics[width=1\linewidth]{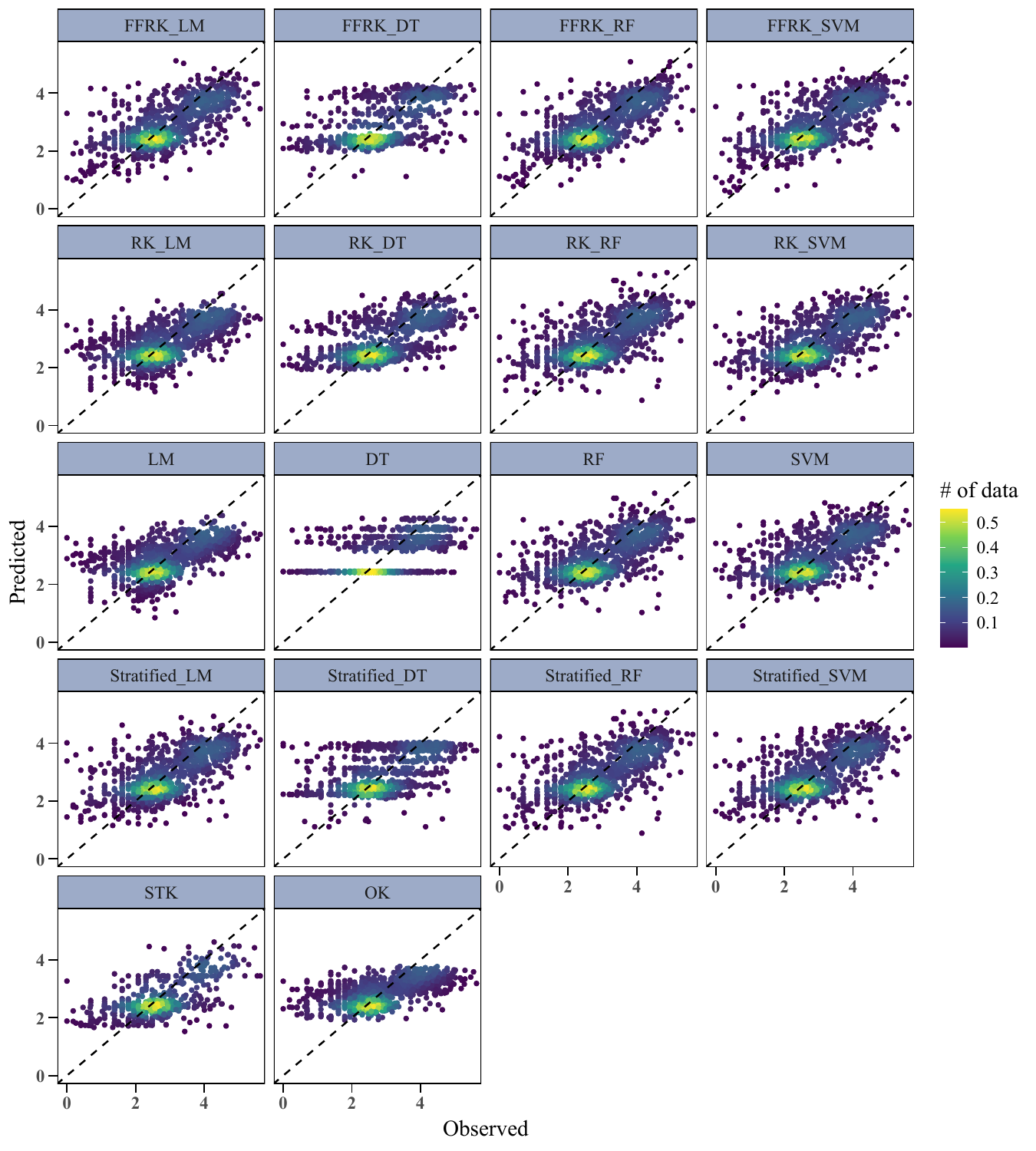}
    \caption{Scatter plot of the prediction results for Cu}
    \label{fig:cu}
\end{figure*}

\begin{figure*}[htbp]
    \centering
    \includegraphics[width=1\linewidth]{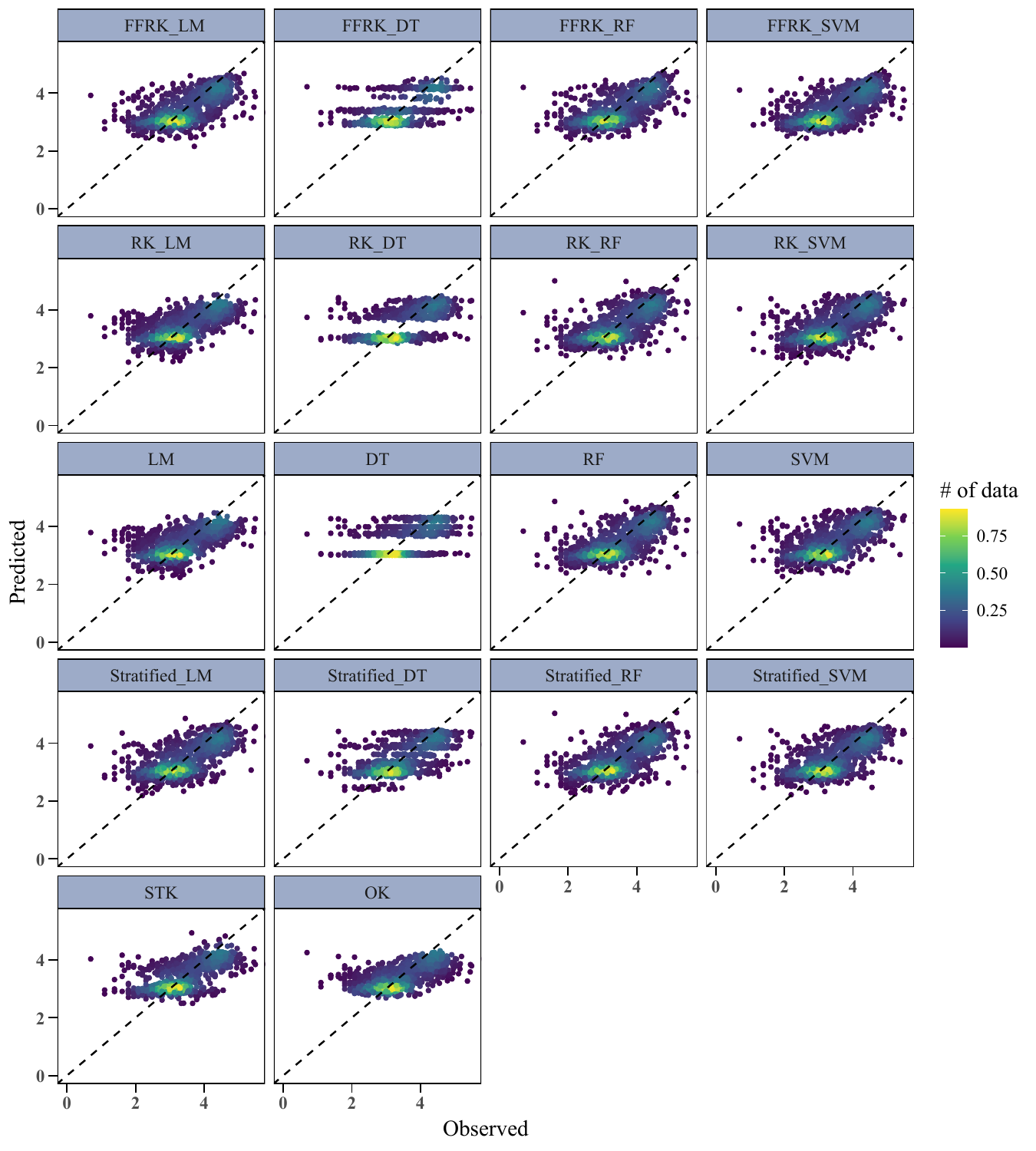}
    \caption{Scatter plot of the prediction results for Zn}
    \label{fig:zn}
\end{figure*}

\begin{figure*}[htbp]
    \centering
    \includegraphics[width=1\linewidth]{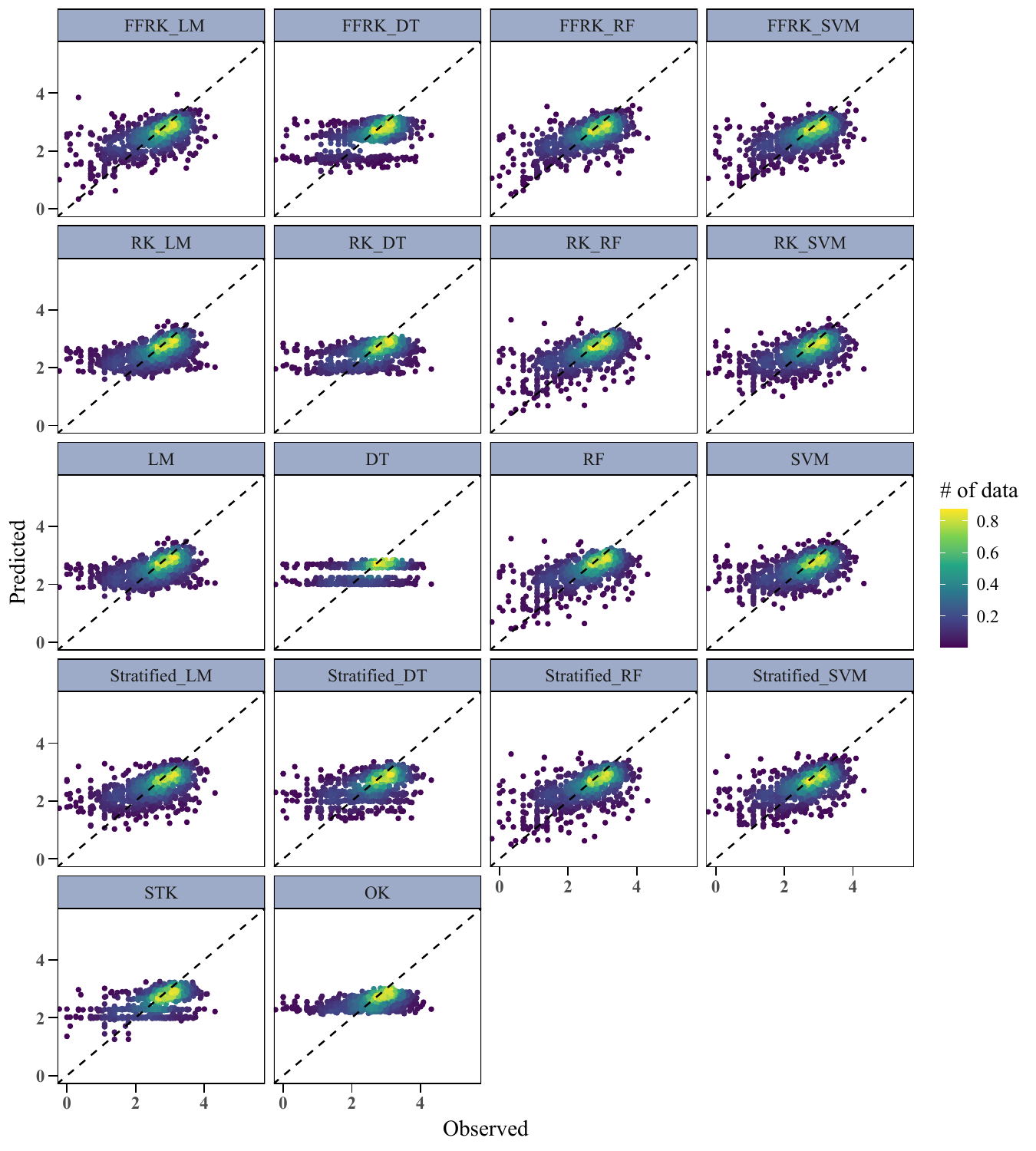}
    \caption{Scatter plot of the prediction results for Pb}
    \label{fig:pb}
\end{figure*}

\end{document}